\documentclass[
twocolumn,
aps,prd,
nofootinbib,
superscriptaddress,
tightenlines,
amsmath,
amssymb
]{revtex4}
\usepackage{bm}
\usepackage{color,graphicx}
\usepackage{amsmath}
\usepackage{amssymb}

\begin{document}
\title{Bounds on the Equation of State of Neutron Stars from High Energy Deeply Virtual Exclusive Experiments} 

\author{Abha Rajan} 
\email{arajan@bnl.gov}
\affiliation{Physics Department, Brookhaven National Laboratory, Upton, New York 11973, USA.}

\author{Tyler Gorda} 
\email{tdg5cs@virginia.edu}
\affiliation{Department of Physics, University of Virginia, Charlottesville, VA 22904, USA.}

\author{Simonetta Liuti} 
\email{sl4y@virginia.edu}
\affiliation{Department of Physics, University of Virginia, Charlottesville, VA 22904, USA.}

\author{Kent Yagi} 
\email{ky5t@virginia.edu}
\affiliation{Department of Physics, University of Virginia, Charlottesville, VA 22904, USA.}

\begin{abstract}
The recent detection of gravitational waves from merging neutron star events has opened a new window on the many unknown aspects of their internal dynamics. A key role in this context is played by the transition from baryon to quark matter described in the neutron star  equation of state (EoS). In particular, the binary pulsar observation of heavy neutron stars 
requires appropriately stiff dense matter in order to counter gravitational collapse, at variance with the predictions of many phenomenological quark models. 
On the other side, the LIGO observations favor a softer EoS therefore providing a lower bound to the equation stiffness. 
We introduce a quantum chromodynamics (QCD) description of the neutron star's high baryon density regime where the pressure and energy density distributions are directly obtained from the matrix elements of the QCD energy momentum tensor. 
Recent ab initio calculations allow us to evaluate the energy-momentum tensor in a model independent way including both quark and gluon degrees of freedom. 
Our approach is a first effort to replace quark models and effective gluon interactions with a first principles, fully QCD-based description. Most importantly, the QCD energy momentum tensor matrix elements are connected to the Mellin moments of the generalized parton distributions which can be measured in deeply virtual exclusive scattering experiments. 
As a consequence, we establish a connection between observables from high energy experiments and from the analysis of gravitational wave events. Both can be used to mutually constrain the respective sets of data.     
In particular, the emerging QCD-based picture is consistent with the GW170817 neutron star merger event once we allow a first-order phase transition from a low-density nuclear matter EoS to the newly-constructed high-density quark-gluon one. 
\end{abstract}

\maketitle
\baselineskip 3.0ex
The Gravitational Wave (GW) observation of a binary neutron star merger \cite{TheLIGOScientific:2017qsa}, GW170817, 
has impacted profoundly the study of the strong interactions by providing for the first time a direct experimental access to nuclear matter at the highest known densities. 
Several theoretical scenarios have been proposed for the nature of the high density regime of strongly interacting systems whose predictions can be summarized in the nuclear matter Equation of State (EoS) relating pressure and energy density. The degrees of freedom in the EoS range, with increasing density, from protons, neutrons and either hyperons, or kaon condensates, to quark matter.
%
%
Neutron stars, formed after deaths of stars more massive than the Sun, offer a natural testbed to probe the microscopic composition of dense nuclear matter. 
Typical neutron stars have masses that are comparable to that of the Sun, and yet the radius is only $\sim 12$km. Due to this extreme compactness, the central density of neutron stars can easily exceed the nuclear saturation density.

A stringent constraint on the EoS is obtained from the observation of $2M_\odot$ pulsars \cite{1.97NS,2.01NS}, requiring it to be stiff, consistently with the predictions for ordinary nuclear matter composed of mostly neutrons and few protons undergoing two and three body interactions.
Understanding the lack of hyperons or kaon condensates is at the center of an intense research effort bringing once more to the forefront the question of the role and size of strange matter in hadron structure.  
More recently, GW170817 has provided an additional bound originating from the measurement of the Tidal Deformability (TD) of compact stars \cite{TheLIGOScientific:2017qsa,De:2018uhw,Most:2018hfd,Abbott:2018wiz}. As the two neutron stars in a binary come closer together due to the loss of binding energy through gravitational wave emission, one of the neutron stars is tidally deformed by the tidal gravitational field created by its companion neutron star. The amount of such tidal deformation is controlled by TD, which depends on the dense matter EoS. 

%

Nuclear (hadronic) and quark matter are generally described as two distinct phases which govern the EoS for a given baryon number density. 
In the quark matter sector, in particular, a variety of models  have been developed throughout the years and are currently used for constructing neutron stars EoSs consistent with current phenomenology (for a  recent review see {\it e.g.}  Ref.\cite{Baym:2017whm}). 

In this Letter we propose a new 
way of evaluating the EoS in the quark matter phase by inferring it directly from the matrix elements of the QCD Energy Momentum Tensor (EMT) between nucleon states. The latter have been evaluated in a series of lattice QCD calculations for quarks in Refs.\cite{Yang:2018nqn,Alexandrou:2016ekb,Deka:2013zha,Hagler:2009ni,Hagler:2007xi} and for gluons in Ref.\cite{Shanahan:2018pib}, 
{whereas a large part of the experimental program at Jefferson Lab @12 GeV is dedicated to extracting them from deeply virtual exclusive electron scattering experiments \cite{dHose:2016mda,Kumericki:2016ehc}.
These matrix elements can be Fourier transformed to give us the energy/momentum, angular momentum and pressure spatial distributions of quarks and gluons.}

{Our description of neutron stars through the QCD EMT is in essence a local density approximation 
motivated by the fact that quark-gluon interactions are of
relatively short range whereby the dominant effects occur
near the center of momentum \cite{PhysRevC.1.1260}. In this regime the interactions between partons from different nucleons are subleading.}

The QCD EMT, stemming directly from the QCD Lagrangian, is defined as,
\begin{equation}
T^{\mu\nu}_{QCD} = \frac{1}{4} \,\,  \overline{\psi} \, \gamma^{(\mu} D^{\nu)} \psi  + Tr \left\{ F^{\mu\alpha} F_\alpha^\nu - \frac{1}{2} g^{\mu \nu}F^2\right\},
\end{equation}
where $\psi$ and $F^{\mu\alpha}$ are the quark and gluon fields, respectively, while $g^{\mu\nu}$ is the spacetime metric.
The EMT matrix element between nucleon states was parametrized in Ref.\cite{Ji:1996nm} as,
\begin{widetext}
\begin{eqnarray}
\langle p', s' \! \mid T_{q,g}^{\mu \nu} \mid \! p, s \rangle = \bar{U}_{s'}(p') \left[A_{q,g}(t) \gamma^{(\mu} P^{\nu)} + B_{q,g}(t) \frac{P^{( \mu} i \sigma^{\nu ) \rho} \Delta_\rho}{2M}  
+ \frac{1}{4M} C_{q,g}(t)\left(\Delta^\mu \Delta^\nu - g^{\mu \nu} \Delta^2\right) + \overline{C}_{q,g}(t) g^{\mu \nu} M \right] U_s(p),
\label{eq:emt_param}
\end{eqnarray}
\end{widetext}
where $q$ and $g$ are the quark and gluon labels; $M$ is the nucleon mass; the initial (final) nucleon spinor is $U_s(p)$ ($\bar{U}_{s'}(p')$); 
$P=(p+p')/2$, and the momentum transfer is, $\Delta=p'-p$, $t=\Delta^2<0$.
The EMT time components encode the densities and flux densities of the quark and gluon fields energy and momentum so that by integrating over the volume element, $d^3 x$, and summing over the quark and gluon components, one obtains the system's total energy, $T^{00}$, and momentum, $T^{0i}$, $(i=1,2,3)$, respectively. 
Similarly, following the structure of the mechanical EMT, the $T^{ij}$ elements can be identified with the pressure ($i=j$), and the shear forces ($i\neq j$). Evaluating the matrix elements from the parametrization in Eq.(\ref{eq:emt_param}) one finds that $A_{q,g}(t=0)$ represents the total quark/gluon momentum relative to the nucleon momentum. The form factor, $A_{q,g}(t)$, therefore provides information on how the momentum is spatially distributed inside the nucleon. Similarly, $t C_{q,g}(t)$ connects to the spatial distribution of pressure inside the nucleon \cite{Polyakov:2002yz,Polyakov:2002wz,Polyakov:2018zvc,Lorce:2018egm}.
\footnote{The combination $A_{q,g}(t)+B_{q,g}(t)$ gives the proton angular momentum. Although measuring this quantity is a major quest for solving the proton spin crisis and we will not address it in this paper.}

The $T_{00}$ element allows us to describe the contribution of the quarks and gluons to the proton mass \cite{Ji:1994av}. Although the quark and gluon terms are separately renormalization scale dependent, their sum leads to conserved quantities and it is therefore apt to represent the mechanical properties of hadronic matter. In order to connect with various quark matter models described in the literature \cite{Baym:2017whm}, we notice that our description holds in the short distance/high density regime, where the nucleon can be considered a statistically large system. 

Through the Operator Product Expansion (OPE) in QCD one connects the EMT form factors,  $A(t), B(t), C(t)$, in Eq.(\ref{eq:emt_param}) with the matrix elements of local twist two operators \cite{Diehl:2003ny,Belitsky:2005qn}. 
The latter are identified with the second Mellin moments of the 
 Generalized Parton Distributions (GPDs) \cite{Ji:1996ek,Ji:1996nm}. 
GPDs were first introduced to define the quark and gluon angular momentum in QCD in terms of observables from lepton proton scattering experiments. They enter, specifically, the hadronic matrix elements for deeply virtual Compton scattering off a nucleon with momentum, $p$, namely the process: $e p \rightarrow e' p' \gamma$, $\gamma$ being a real photon, and its related channels  (see reviews in \cite{Diehl:2003ny,Belitsky:2005qn,Kumericki:2016ehc}). 
In particular, labeling the GPD moments as $A^{q,g}_{2}(t)$, $B^{q,g}_{2}(t)$, $C^{q,g}_{2}(t)$, one has the following relations,
\begin{equation}
A_{q,g}(t) = A^{q,g}_{2}(t), \;\;   B_{q,g}(t) = B^{q,g}_2(t),  \;\; C_{q,g}(t) = 4 C^{q,g}_{2}(t).
\end{equation} 
The combination of GPD moments $A_{2}^{q,g}+B_{2}^{q,g}$ was extracted in Refs.\cite{Mazouz:2007aa,Ye:2006gza}, although model dependently. More recently, in Ref.\cite{Burkert:2018bqq}, it was possible to determine $C_{2}^q(t)$, opening a very first window on the pressure distribution inside the proton. More precise and copious determinations are on their way within the experimental program of Jefferson Lab at 12 GeV and the planned future Electron Ion Collider (EIC) \cite{Accardi:2012qut}.
It should be noted that the GPDs also depend on the renormalization scale in such a way that the sum over the quark and gluon terms is scale independent. The scale chosen to evaluate separately the quark and gluon terms in lattice QCD is $\mu^2=4$ GeV$^2$ \cite{Shanahan:2018pib,Yang:2018nqn,Alexandrou:2016ekb,Deka:2013zha,Hagler:2009ni,Hagler:2007xi}.

In order to define the energy and pressure spatial distributions, $\varepsilon(r)$, and $p(r)$, we first  
introduce probability density distribution, $\rho_{\Lambda\lambda}({\bf b})$ to find a quark with helicity $\lambda$ located inside the nucleon (with with helicity $\Lambda$) at position ${\bf b}$ from the nucleon's center of momentum, in the transverse plane \cite{Soper:1976jc,Burkardt:2000za,Ralston:2001xs,Diehl:2002he,Belitsky:2003nz,Liuti:2004hd,Lorce:2011kd}
\begin{eqnarray}
\rho^{q}_{\Lambda\lambda}({\bf b}) = H_{q}({\bf b}^2) + \frac{b^i}{M} \epsilon_{ij} S_T^j \frac{\partial}{\partial b} E_{q}({\bf b}^2) +  \Lambda \lambda \widetilde{H}_{q}({\bf b}^2),
\end{eqnarray}
where $i=1,2$, and $S_T^i$ is the transverse proton spin. $H_{q}({\bf b}^2), E_{q}({\bf b}^2)$, $\widetilde{H}_{q}({\bf b}^2)$,  are the Fourier transforms in the transverse plane of the $t$-dependent GPDs corresponding to different quark-proton polarization configurations. $t$, the four-momentum transfer squared introduce previously, is related to the transverse momentum transfer, ${\bf \Delta}_T$ as: $t = t_0 - {\Delta_T^2}/({1-\xi^2})$, where $t_0 = - {4 \xi^2 M^2}/({1-\xi^2})$, and $\xi$ is a longitudinal momentum fraction.
For an unpolarized quark in an unpolarized proton we have,
\begin{eqnarray}
\sum_{\Lambda,\lambda} \rho^{q}_{\Lambda\lambda}({\bf b}) = H_q({\bf b}^2)=
\int \frac{d^2 {\bf \Delta}_T}{(2\pi)^2} e^{i{\bf \Delta}_T \cdot \bf b} A^{q}_1(t),
\end{eqnarray}
where $A^{q}_{1}(t)$, is the quark contribution to the nucleon Dirac form factor. 
\begin{figure}
\begin{center}
\includegraphics[width=8.5cm]{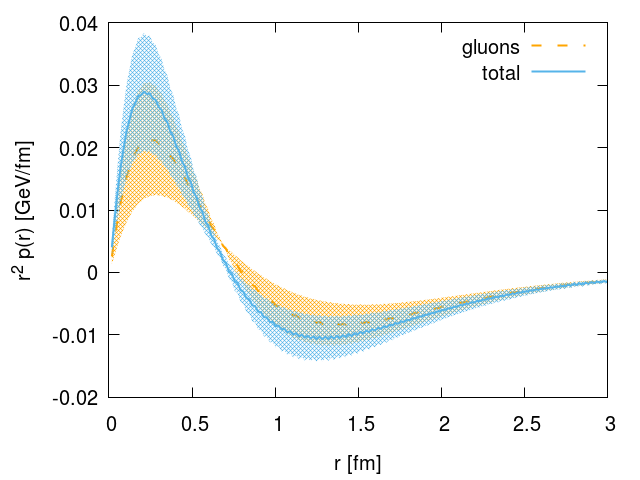}
\caption{The pressure as a function of radial distance, obtained using $C_{2,q}(t)$ and $C_{2,g}(t)$. We use the dipole form for $C_{2,g}(t)$ taken from Ref.\cite{Shanahan:2018pib}. The quark contribution is obtained by fitting a dipole form to data obtained from Ref.\cite{Hagler:2007xi} and \cite{Martha}. The shaded area is the error  obtained from the parameters of the fits to the lattice QCD data.}
\end{center}
\label{fig:FF}
\end{figure}
Similarly, denoting $\mid {\bf b} \mid = r$, we define the energy density and pressure distributions over the transverse plane, $\varepsilon (r)$, and $p(r)$, respectively, as the Fourier transforms of $A_2^{q,g}(t)$ and $ t\, C_2^{q,g}(t)$,
\begin{eqnarray}
\label{energy}
\varepsilon_{q,g}(r) & = & \!\! 
 \int \frac{d^2 {\bf \Delta}_T}{(2\pi)^2} e^{i {\bf \Delta}_T \cdot \bf b} A^{q,g}_2(t), \\
\label{pressure} 
p_{q,g}(r) & = & \!\!
\int \frac{d^2 {\bf \Delta}_T}{(2\pi)^2} e^{i{\bf \Delta}_T \cdot \bf b} \, \frac{t}{M^2} \, C^{q,g}_2(t).
\end{eqnarray}
The total energy density and pressure distribution are obtained as the sum of the quark flavor singlet and gluon terms.
The Fourier transforms were performed using the FFTW package \cite{Frigo:2004:FFT}. As the form factors $A_2^{q,g}(t)$ and $C_2^{q,g}(t)$  are symmetric in the azimuthal angle $\phi_{\Delta_T}$ or, in other words, symmetric in $\Delta_x$ and $\Delta_y$, their Fourier transforms are purely real, they have only radial dependence, and they can be therefore extrapolated to describe 3D configurations. 
{To accomplish this we used the Abel transform \cite{Moiseeva:2008qd}.}
\footnote{
{Note that, implicit in the Abel transform is another unit of length, and multiplying $A$ and $C$ by a factor $M$ allows us to define the 3D energy density and pressure.}}

The Fourier transforms in the gluon sector were calculated from the lattice QCD evaluations of Ref.\cite{Shanahan:2018pib}. The quark isoscalar combination, $u+d$, was obtained Fourier transforming the lattice QCD results of both Refs.\cite{Hagler:2007xi} and \cite{Martha}. In both the gluon and quark case the given range of $t$ values is not sufficiently large to allow a precise Fourier transformation. We therefore used the dipole form, $a/(1-t/b^2)^2$, to fit the data on the form factors. Not only does this allow us to have a smooth fall off at large $t$, in the case of $C_2^{q,g}$ it also allows us to extrapolate to $t$ close to zero where there are relatively few data points with large uncertainties. The error on the fit parameters is the main source of error in the pressure and energy density distributions that are obtained after the Fourier transform.

We can now make the connection between the energy density and pressure for quark gluon matter, respectively, and neutron stars. To construct the solution for the latter in General Relativity, one needs to solve the Einstein equations, that state how the spacetime is curved for a given matter distribution. To be more precise, the energy momentum tensor, controlled by the matter energy density and pressure, determines the curvature of spacetime. 

Our main result is that the EoS obtained from the EMT is dominated by the gluon contribution, the quark contribution being largely suppressed. 
%
\begin{figure}
\begin{center}
\includegraphics[width=8.5 cm]{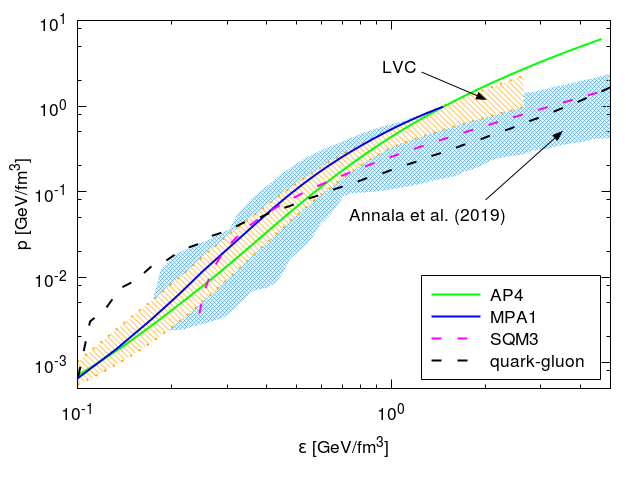}
\caption{Comparison of various EoSs. We present the quark-gluon EoS (black dashed) constructed in this letter, together with two hadronic EoS AP4 (green solid) and MPA1 (blue solid) and one quark matter EoS SQM3 (magenta dashed). The orange shaded region (LVC) is the allowed region from GW170817. The blue shaded region represents the family of all possible NS-matter EoSs, obtained
with the speed-of-sound interpolation method introduced in Ref.\cite{Annala:2019puf}.
}
\label{fig:EoS}
\end{center}
\end{figure}
\begin{figure}
\begin{center}
\includegraphics[width=8.5cm]{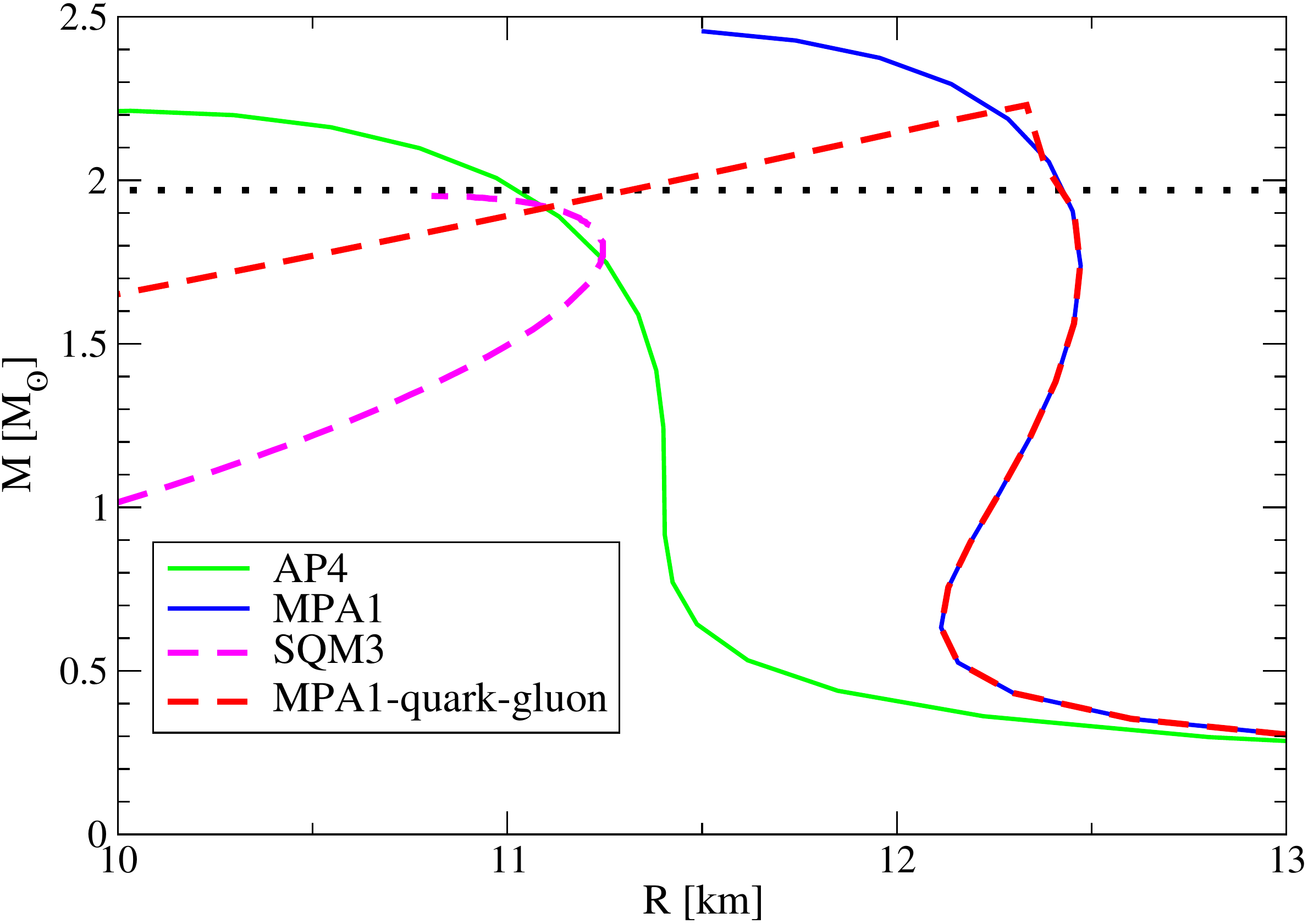}
\caption{Mass-radius relation for NSs with various EoSs. The quark-gluon EoS is stitched to hadronic MPA1 EoS. The horizontal line at $1.97M_\odot$ corresponds to the lower bound on the mass of the pulsar J0348+0432 \cite{2.01NS}.}
\label{fig:MR}
\end{center}
\end{figure}
We eliminate $r$ between $\varepsilon$ and $p$ in Eqs.(\ref{energy},\ref{pressure}), then we plot the QCD values and we compare with previous EoSs. Figure \ref{fig:EoS} shows the  quark-gluon EoS constructed here. For reference, in Fig. \ref{fig:EoS} we also present EoSs for two hadronic matter (AP4 \cite{Akmal:1998cf} and MPA1 \cite{Muther:1987xaa}) and one quark matter (SQM3 \cite{Lattimer:2000nx}). The latter is a strange quark matter EoS based on 
an MIT-type bag model including QCD interactions and the strange quark mass term. Observe that the new quark-gluon EoS is very similar to SQM3 in the range $0.5< \varepsilon < 2$ GeV/fm$^3$. 
We also show the range constrained by LIGO with GW170817 \cite{Abbott:2018exr}, though this range should only be taken as a guidance since (i) LIGO placed bounds on the pressure vs mass density plane, not the pressure vs energy density plane (we converted the former to the latter using the first law of thermodynamics), and (ii) the constraint is obtained assuming that the EoS is smooth and continuous (and thus does not necessarily apply to EoSs with hadron-quark phase transitions.)

Figure \ref{fig:MR} presents the mass-radius relation for neutron stars. If we use our quark-gluon EoS, we find that the stellar radius is larger than 25km, which is not observationally favored. A more realistic EoS can be constructed by stitching the quark-gluon EoS to a hadronic EoS (see e.g. \cite{Paschalidis:2017qmb,Montana:2018bkb} for probing such hybrid EoSs with GW170817). To give an example, we stitched the former to MPA1 which is consistent with the GW170817 constraint. We chose the transition pressure to be 0.2 GeV/fm$^3$ so that the maximum mass of a hybrid star (quark matter core with hadronic matter envelope) exceeds 1.97$M_\odot$ that corresponds to the lower bound on the mass of the pulsar J0348+0432 \cite{2.01NS}. Thus, the new EoS constructed here is consistent with NS observations. 

In conclusion, we made a connection between the pressure and energy density in neutron stars and yet another set of collider observables, the GPDs. 
The most important implication of our work is that the EoS of dense matter in QCD can be obtained from first principles, using ab initio calculations for both quark and gluon degrees of freedom. Gluons, in particular, dominate the EoS, and provide a trend in the high density regime which is consistent with the constraint from LIGO.
The proposed line of research opens up a new framework for understanding the properties of hybrid stars. In the future we hope to set more stringent constraints on the current controversy about the nature of the hadron to quark matter transition at zero temperature.

We thank M. Constantinou for providing recent lattice evaluations in the quark sector from the European
Twisted Mass Collaboration, and G. Baym, M. Engelhardt, C. Lorc\'e, M. Polyakov for useful discussions and comments. This research was funded by DOE grants DE-SC0016286 (S.L. and A.R.), DE-SC0012704 (A.R.), DE-SC0007984 (TG), NSF Award PHY-1806776 (K.Y.). K.Y. would like to also acknowledge networking support by the COST Action GWverse CA16104; A.R. acknowledges the LDRD grant from Brookhaven Science Associates.



\bibliography{OAM_bib}
\end{document}